\documentclass[%
superscriptaddress,
notitlepage,
showpacs,
amsmath,amssymb,
aps,
onecolumn,
longbibliography,
]{revtex4-1}

\usepackage{psfrag,graphicx,epsfig,color}% Include figure files
\usepackage{dcolumn}% Align table columns on decimal point
\usepackage{bm}% bold math
\usepackage{natbib}
\usepackage{float}
\usepackage[usenames,dvipsnames,svgnames,table]{xcolor}
\usepackage{subfigure}
\usepackage{nicefrac}

\graphicspath{{figs/}{plots/}}

\def\re    {{R_\lambda}}
\def\uu {{\mathbf{u}}}
\def\xx {{\mathbf{x}}}

\definecolor{mygreen}{rgb}{0,0.7,0.}

\begin{document}

\title{
Comparing velocity and passive scalar statistics in fluid turbulence \\ at high Schmidt numbers and Reynolds numbers
}

%%% Authors %%%
\author{Dhawal Buaria }
\email[]{dhawal.buaria@nyu.edu}
%\thanks{}
\affiliation{Tandon School of Engineering, New York University, New York, NY 11201, USA}
\affiliation{Max Planck Institute for Dynamics and Self-Organization, 37077 G\"ottingen, Germany}
\author{Katepalli R. Sreenivasan}
\affiliation{Tandon School of Engineering, New York University, New York, NY 11201, USA}
\affiliation{Department of Physics and the Courant Institute of Mathematical Sciences,
New York University, New York, NY 10012, USA}

\date{\today}% It is always \today, today,
             %  but any date may be explicitly specified

\begin{abstract}

Recently, Shete et al. [Phys. Rev. Fluids 7, 024601 (2022)]
explored the characteristics of passive scalars
in the presence of a uniform mean gradient, mixed by stationary isotropic turbulence. 
They concluded that at high Reynolds and Schmidt numbers, the presence
of both inertial-convective and viscous-convective ranges,
renders the statistics of the scalar and velocity 
fluctuations to behave similarly. However, their data included Schmidt 
numbers of 0.1, 0.7, 1.0 and 7.0, only the last of which can 
(at best) be regarded as moderately high. Additionally, they do not
consider already available data in the literature at substantially 
higher Schmidt number of up to 512. By including these data, we demonstrate here
that the differences between velocity and scalar
statistics show no vanishing trends with increasing Reynolds and Schmidt numbers, 
and essential differences remain in tact at all Reynolds and Schmidt numbers.

%The analysis of Shete et al. is limited
%to just a single data point of $Sc=7$ (in the
%high Schmidt number range), which 
%is insufficient to draw any definitive conclusions,
%especially in the high $Sc$ limit.

\end{abstract}

\maketitle

%################################################

In Ref.~\cite{shete2022}, Shete et al. 
investigate the mixing of a passive
scalar $\theta (\xx,t)$ in isotropic turbulence, driven by a uniform mean-gradient
$\nabla \Theta = (G, 0, 0)$:
\begin{align}
\partial \theta /\partial t + \uu \cdot \nabla \theta
= - \uu \cdot \nabla \Theta + D \nabla^2 \theta,
\end{align}
where $D$ is the scalar diffusivity and 
$\uu(\xx,t)$ is the underlying
turbulent velocity field governed by the
incompressible Navier-Stokes equations.
The mixing characteristics are
governed by two parameters: the Schmidt number
$Sc=\nu/D$, $\nu$ being the kinematic
viscosity of the fluid
and the Taylor-scale Reynolds number 
$\re=u^\prime \lambda/\nu$, where
$u^\prime$ is the root-mean-square velocity
fluctuation and $\lambda$ is the Taylor length scale. 
The data of Ref.~\cite{shete2022},
obtained from state-of-the-art direct numerical simulations (DNS), 
correspond to 
$\re=633$ and $Sc=0.1, 0.7, 1.0, 7.0$. The Reynolds number is high enough 
to display inertial range characteristics, but the Schmidt number 
range is very limited. 
Given that turbulent mixing for $Sc>1$ is fundamentally different from
that for $Sc<1$, the authors essentially have a single data point 
at $Sc = 7$ in the $Sc>1$ regime, making their inferences unsound.
Further, the authors ignored the data already 
available in the literature at 
much higher Schmidt numbers. 
Here, we include them as well and 
demonstrate that the conclusions in \cite{shete2022}
are not correct and need to be revised. 

The analysis in Ref.~\cite{shete2022} 
is built around the following three points: 
(1) the skewness of scalar gradients; 
(2) a comparison of the intermittency exponent of scalar dissipaton rate to that of energy dissipation rate; 
and (3) a comparison between the probability density functions 
(PDFs) of scalar and energy dissipation rates. 
Note that for an eddy of characteristic size $r$, 
inertial and inertial-convective ranges are both
defined by the condition $\eta_K \ll r \ll L$, where
$\eta_K$ is the Kolmogorov length scale, marking the
viscous cutoff, and $L$ is the 
large scale at which the energy is injected;
the viscous convective range is defined by
$\eta_B \ll r \ll \eta_K$, where
$\eta_B = \eta_K Sc^{-1/2}$ is the Batchelor length scale.
Evidently, a fully developed viscous convective range
requires $Sc \gg 1$ (and unlikely to exist at $Sc = 7$). 
Here, we assess each of the three points above
and show that fundamental differences remain between velocity and 
scalar statistics even at high $Sc$, 
essentially demonstrating -- contrary to the conclusion of \cite{shete2022} -- 
that velocity and scalar statistics
are never similar. 
%This conclusion does not come as a surprise, 
%but principally sets the record straight.

\subsection{Anisotropy of scalar gradients}

It is now well known that local isotropy is 
violated  for a passive scalar
driven by a uniform mean-gradient \cite{sab_79, ZW00, sreeni19}. 
Specifically, for the 
scalar gradient in the direction
of the imposed mean-gradient,
$\nabla_\| \theta$, the odd-moments 
are non-zero of the order unity
(whereas local isotropy requires them to be zero). 
This violation of local isotropy can be traced to the existence of so-called
ramp-cliff structures \cite{sab_79, sreeni19, BCSY21b}, 
resulting from a direct influence of the imposed large-scale mean gradient
on the small-scale scalar field  
\footnote{In fact, even in absence of a large-scale mean gradient, similar structures
are observed in the scalar field \cite{watanabe2004}}.
Based on the specific ramp-cliff model of \cite{sreeni19},
Shete et al.\ reported the following expression
(Eq.~10 of ref.~\cite{shete2022}):
\begin{align}
\frac{ \langle (\nabla_\| \theta )^p  \rangle}{ \langle (\nabla_\| \theta)^2  \rangle^{p/2}}
\sim Sc^{-1/2} R_\lambda^{(p-3)/2} \ , \ \ \ \ \ 
p=3,5,7...,
\label{eq:odd1}
\end{align}
which quantifies the scaling of odd-moments of $\nabla_\| \theta$.
Using only two data points for $Sc=1,7$ 
and restricting the testing for $p=3$ 
(see Fig.~7 of Ref.~\cite{shete2022}), the authors concluded that 
Eq.~\eqref{eq:odd1} is valid.

We first note that the result in Eq.~\eqref{eq:odd1}, 
implicitly given in Ref.~\cite{sreeni19}, was  explicitly
derived in \cite{BCSY21b}, which Shete et al.
did not recognize. 
Additionally, the authors also ignore that Eq.~\eqref{eq:odd1},
along with the underlying assumptions, was rigorously tested 
in Ref.~\cite{BCSY21b} by
using DNS data over a large range of 
Schmidt numbers $Sc=1-512$ (at $\re=140$) and also for moment orders $p = 3, 5$ and 7. 
Comprehensive details about the DNS and numerical methods
are available in \cite{clay.cpc1, clay.cpc2, clay.omp}, whereas
the database along with simulation parameters is outlined in \cite{BCSY21a, BCSY21b}.
In Ref.~\cite{BCSY21b}, it was found that the original ramp-cliff model required modifications 
and the scaling of odd moments of $\nabla_\| \theta$ is better
described by the following expression:
\begin{align}
\frac{ \langle (\nabla_\| \theta )^p  \rangle}{ \langle (\nabla_\| \theta)^2  \rangle^{p/2}}
\sim Sc^{-1/2 + \alpha} \ R_\lambda^{(p-3)/2} \ , 
\ \ \ \ \ \ \  \text{for} \ \ p=3,5,7... \ ,
\label{eq:odd2}
\end{align}
where the new exponent on $Sc$, with $\alpha \approx 0.05$, represents
a slightly weaker slope compared to $-1/2$ in Eq.~(\ref{eq:odd1}). 
As shown in \cite{BCSY21b}, this implies that a new quantity 
$\eta_D = \eta_B Sc^{\alpha}$ ($\alpha \approx 0.05$)
marks the true diffusive cutoff scale in the scalar
field (instead of $\eta_B$). 
The difference between $\eta_B$ and $\eta_D$ arises because 
the scalar dissipation anomaly does not hold for large
$Sc$ \cite{BSXDY, BYS.2016, BCSY21a}. A similar idea was also
proposed in an independent study \cite{yasuda20}.

%and found the need for the following modification:
%, whereby
%the scalar gradient at the %cliff 
%is assumed to be $\theta_{\rm rms}/\eta_B$,
%where $\theta_{\rm rms}$ is %the rms of
%scalar fluctuations.
%Following this assumption, %it 
%was shown in ref.~\cite{BCSY21b},
%that when the odd moments of $\nabla_\| \theta$ 
%are appropriately normalized by $\theta_{\rm rms}/\eta_B$, 
%the following result is %obtained: 
%\begin{align}
%\frac{ \langle (\nabla_\| %\theta )^p  %\rangle}{(\theta_{\rm %rms}/\eta_B)^p }
%\sim Sc^{-1/2} R_\lambda^{-3/2} %\ , \ \ \ \ \ 
%p=3,5,7..
%\label{eq:oddrms}
%\end{align}
%which has the same $Sc^{-1/2}$ dependence
%as Eq.~\eqref{eq:odd1}. However,
%since $\theta_{rms}/\eta_B$ depends on $Sc$,
%increasing deviations from $-1/2$ scaling appear in Eq.~Eq.~\eqref{eq:odd1} as $p$ increases
%(see Fig.~4 of \cite{BCSY21b}). Available data suggest that $\theta_{\rm rms}/\eta_D$,
%It follows that 
%the odd moments
%of $\nabla_\| \theta$ are then given as 
%\begin{align}
%\nonumber
%\frac{ \langle (\nabla_\| \theta %)^p  \rangle}{(\theta_{\rm %rms}/\eta_B)^p }
%\sim Sc^{\beta_p} %R_\lambda^{-3/2} \ , \ \ \ \ \ 
%p=3,5,7..     \\ 
%\ \ \ \ \text{with} \ \ \ %\beta_p = -1/2 + \alpha - \alpha p 
%\label{eq:oddrms2}
%\end{align}
%the normalized moments obey: 

\begin{figure}
\begin{center}
\includegraphics[width=0.5\linewidth]{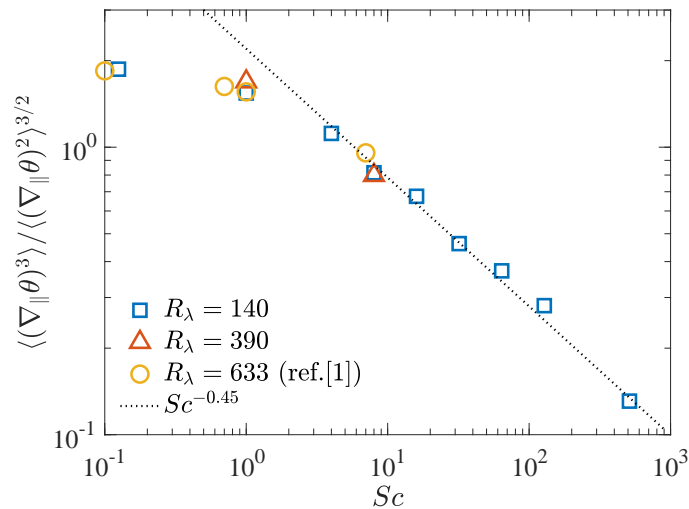}
\caption{
Skewness of $\nabla_\| \theta$ as a function
of $Sc$, for various
$\re$. The data for $\re=390$ and $633$ are
respectively adjusted by factors 1.2 and 0.9 
to account for a very weak $\re$ dependence,
but do not have any bearing on trends with $Sc$.
}
\label{fig:skew}
\end{center}
\end{figure}

%The modification in \eqref{eq:odd2} 
%has been validated for $p=3,5,7$ in \cite{BCSY21b}, going up to 
%very high $Sc=512$. 
%However, they were not %considered
%by Shete et al. in %ref.~\cite{shete2022}. 

A reinspection of Fig. 7 of ~\cite{shete2022} shows that
the data for $Sc=1,7$  noticeably depart from the
$Sc^{-1/2}$ scaling but are consistent with the updated
result in Eq.~\eqref{eq:odd2}.
For completeness, we combine the data for $p=3$ 
for various $\re$ and $Sc$ in
Fig.~\ref{fig:skew} (including 
previously unreported data at $\re=390$,
which is also large enough for inertial-range scaling 
to exist \cite{BCSY21a}). 
Evidently, the data
at $\re=140$ are consistent with modified
$Sc$ scaling in Eq.~\eqref{eq:odd2} (with $\alpha=0.05$).
Data at higher $\re$ are also in agreement with 
the trends at $\re=140$, though significantly higher
$Sc$ would be required to see if $\alpha$ has a  $\re$-dependence. 
However, Shete et al. did not examine the results for $p=5,7$, nor 
tested the veracity of the length scale
$\eta_D$, both of which are especially important given the lack
of high $Sc$ data in their work. 

Notwithstanding the quantitative differences, it is clear that
the odd moments
decrease as $Sc$ increases, presumably approaching zero, i.e., local isotropy is 
restored at infinite $Sc$.
Shete et al. invoked this notion to conclude
that due to increased scale separation
at high $\re$ and $Sc$, the scalar statistics
become universal 
and hence similar to velocity statistics. 
This conclusion is unjustified because even though local
isotropy of the scalar is restored at $Sc\to\infty$, scalar dissipation
anomaly is also simultaneously violated \cite{BCSY21a}, while 
the dissipation anomaly for energy dissipation rate
continues to hold. This is a fundamental difference
between velocity and scalar statistics, which implies
that they can never be similar at any $\re$ and $Sc$. Shete et al.
assume that scalar dissipation anomaly is valid
based on just two data points $Sc=1,7$,
-- but also note that this assumption would not hold
at higher $Sc$ (implying that their assumption
and conclusions are contradictory). 

%two main reasons.
%First, the odd moments of $\nabla_\|\theta$
%are strictly zero only at infinite $Sc$, i.e.,
%local isotropy is not achieved at finite $Sc$ 
%and certainly not at $Sc=7$ (which is only marginally
%higher than $Sc=1$). 
%It should also be noted that the odd moments of 
%$\nabla_\|\theta$ also decrease with increasing
%$Sc$ at very low $\re$, at which no inertial 
%range exists; for instance, see
%results of \cite{PK02} at $\re=38$.
%Essentially, recovering local isotropy for scalar gradients
%only requires very high $Sc$ and 
%the presence of (or lack thereof) an inertial range
%has no direct bearing on it. 

\subsection{Intermittency exponent of scalar and velocity gradients}

With respect to the second point, Shete et al. investigated the
intermittency of local  averages (over scale $r$) of energy dissipation
$\epsilon_r$
and scalar dissipation $\chi_r$.
Following Kolmogorov's refined hypothesis \cite{K62},
it is expected that:
\begin{align}
\langle \epsilon_r^2 \rangle \sim r^{-\mu} \ , 
\ \ \ \ 
\langle \chi_r^2 \rangle \sim r^{-\mu_\theta} \ , 
\end{align}
for $r$ in the inertial range (and the inertial-convective range for the scalar). 
Here, $\mu$ is the well-known intermittency exponent for the energy dissipation
and $\mu_\theta$ for the scalar dissipation. 
Previous studies have shown that $\mu \approx 0.25$ and $\mu_\theta \approx 0.35$ for scalars
corresponding to $Sc \sim 1$ \cite{prasad88,sreeni93,MW98}.
In Fig.~8 of Ref.~\cite{shete2022}, Shete et al.  
extract the intermittency
exponents using some approximations and find that, indeed, 
$\mu\approx0.25$, and $\mu_\theta > \mu$ for $Sc \leq1$,
in essential agreement with previous 
studies \cite{prasad88,sreeni93,MW98}. Additionally, 
they infer that $\mu_\theta$ decreases to $0.25$ 
when $Sc = 7$, and conclude once again that 
velocity and scalar statistics
are similar at high $Sc$. 

However, this conclusion is erroneous
because the authors simply captured the slight decline of $\mu_\theta$ at one $Sc$.
Indeed, had they used data from larger $Sc$, it would have been clear that the scalar
intermittency exponents  $\mu_\theta$ 
does not stay matched with $\mu$ for large $Sc$,
but monotonically decreases,
seemingly to zero as $Sc\to\infty$; see Fig.~5 of \cite{BS2022}. Such an approach
to zero is consistent with the violation of scalar dissipation
anomaly at infinite $Sc$. 
Since $\mu_\theta > \mu$ at $Sc=1$
and $\mu_\theta\to0$ at $Sc\to\infty$,
it naturally follows that $\mu_\theta = \mu$ at some intermediate
$Sc$, which is precisely what Shete et al. 
find for $Sc = 7$. But this does not imply
that velocity and scalar statistics
become similar at high $Sc$.

\subsection{PDF of scalar and energy dissipation}

For the third point, Shete et al.  consider the PDFs
of energy dissipation rate $\epsilon$ 
and scalar dissipation rate $\chi$.
Based on their Fig.~9, they conclude that as $Sc$
increases, the PDFs of $\chi$ and $\epsilon$ approach each other. 
Firstly, it can be clearly seen from their Fig.~9
that the two PDFs
do not coincide: the discrepancy
may appear to be small because the plot
shows the PDFs of logarithms of $\epsilon$ and $\chi$.
The differences between the PDFs of $\epsilon$ and $\chi$ would be far more
conspicuous. Furthermore, it is well known that PDFs of highly 
intermittent quantities, such as $\epsilon$ and $\chi$ for $Sc \sim 1$ are close to
log-normal. Thus, it is not a surprise that all the PDFs
shown in Fig.~9 of \cite{shete2022} are close to each other.
However, with much higher range of $Sc$ (albeit at lower $\re$), it has been 
demonstrated previously \cite{schum2005} that PDF of $\chi$
becomes increasingly different from that of $\epsilon$ as $Sc$
increases. Thus the behavior observed by Shete et al.
in Fig.~9 is also just transitory for one $Sc$.

It is worth mentioning that, 
for both the intermittency exponent and PDF,
Shete et al.~utilize the energy dissipation 
to represent small-scale velocity field. However, energy dissipation 
is not the only measure of velocity gradients statistics.
Other measures, such as the enstrophy $\Omega$ 
are equally viable. 
In this regard, it is well known that $\Omega$ is more intermittent that $\epsilon$;
for instance, the intermittency exponent of $\Omega$
is larger than that of $\epsilon$ \cite{chen97,BS2022}.
Moreover, the two PDFs are distinctly different
from each other, even at very high Reynolds numbers
\cite{BPBY2019, BP2022}. Additionally, there 
are numerous other distinct differences
between the fine scale structure of scalar and velocity gradients
\cite{Ashurst87, BBP2020}.
Thus, simply comparing scalar dissipation
with energy dissipation does not attest to the similarity of scalar and velocity statistics.

\subsection{Velocity and scalar structure functions}

In addition to the above three points,  
we consider an additional argument, which was not considered by Shete et al.,
but clearly negates their conclusion. 
The differences between velocity and scalar fields 
can also be demonstrated by comparing their respective structure functions.
If we represent the velocity and scalar increments 
over scale $r$ by 
$\delta_r u$ and $\delta_r \theta$, respectively,
then in the inertial range, the $p$-th order structure
functions are expected to follow the power laws:
\begin{align}
\langle (\delta_r u)^p \rangle \sim r^{\zeta_p} \ , 
\ \ \ \  \text{and} \ \ \ \ 
\langle (\delta_r \theta)^p \rangle \sim r^{\xi_p} \ .
\end{align}
If the velocity and scalar statistics are indeed similar
at high $\re$ and $Sc$, we should obtain $\zeta_p = \xi_p$.
Note, $\zeta_p = \xi_p = p/3$ for the K41
phenomenology, but it is well known that both exponents
strongly depart from K41 due to intermittency, with 
scalar exponents $\xi_p$
departing more strongly \cite{SA97,ZW00}. While
previous studies were mostly restricted to $Sc\sim1$, 
the effect of increasing $Sc$ was more recently studied in \cite{BCSY21a};
where it was found that with increasing $Sc$, the deviation of $\xi_p$
from the K41 result is even stronger. In the limit of $Sc \to \infty$, it was 
observed that $\xi_p$ behave similar to Burgers' turbulence \cite{BCSY21a}. 
Essentially, the discrepancy between $\zeta_p$ and $\xi_p$ {\em increases} with
$Sc$; again demonstrating that velocity and scalar statistics cannot be similar. 

\subsection{Summary and conclusion}

To summarize, by considering the entire data available, we have demonstrated that 
the velocity and scalar fields depart increasingly from each other as the Schmidt 
number increases. This is not a 
surprising conclusion but seems important to set right in view of the erroneous 
conclusion of \cite{shete2022}, based on 
simulations at just one value of $Sc > 1$.

%\bibliography{large_grad,zebib}

\begin{thebibliography}{26}%
\makeatletter
\providecommand \@ifxundefined [1]{%
 \@ifx{#1\undefined}
}%
\providecommand \@ifnum [1]{%
 \ifnum #1\expandafter \@firstoftwo
 \else \expandafter \@secondoftwo
 \fi
}%
\providecommand \@ifx [1]{%
 \ifx #1\expandafter \@firstoftwo
 \else \expandafter \@secondoftwo
 \fi
}%
\providecommand \natexlab [1]{#1}%
\providecommand \enquote  [1]{``#1''}%
\providecommand \bibnamefont  [1]{#1}%
\providecommand \bibfnamefont [1]{#1}%
\providecommand \citenamefont [1]{#1}%
\providecommand \href@noop [0]{\@secondoftwo}%
\providecommand \href [0]{\begingroup \@sanitize@url \@href}%
\providecommand \@href[1]{\@@startlink{#1}\@@href}%
\providecommand \@@href[1]{\endgroup#1\@@endlink}%
\providecommand \@sanitize@url [0]{\catcode `\\12\catcode `\$12\catcode
  `\&12\catcode `\#12\catcode `\^12\catcode `\_12\catcode `\%12\relax}%
\providecommand \@@startlink[1]{}%
\providecommand \@@endlink[0]{}%
\providecommand \url  [0]{\begingroup\@sanitize@url \@url }%
\providecommand \@url [1]{\endgroup\@href {#1}{\urlprefix }}%
\providecommand \urlprefix  [0]{URL }%
\providecommand \Eprint [0]{\href }%
\providecommand \doibase [0]{http://dx.doi.org/}%
\providecommand \selectlanguage [0]{\@gobble}%
\providecommand \bibinfo  [0]{\@secondoftwo}%
\providecommand \bibfield  [0]{\@secondoftwo}%
\providecommand \translation [1]{[#1]}%
\providecommand \BibitemOpen [0]{}%
\providecommand \bibitemStop [0]{}%
\providecommand \bibitemNoStop [0]{.\EOS\space}%
\providecommand \EOS [0]{\spacefactor3000\relax}%
\providecommand \BibitemShut  [1]{\csname bibitem#1\endcsname}%
\let\auto@bib@innerbib\@empty
%</preamble>
\bibitem [{\citenamefont {Shete}\ \emph {et~al.}(2022)\citenamefont {Shete},
  \citenamefont {Boucher}, \citenamefont {Riley},\ and\ \citenamefont
  {de~Bruyn~Kops}}]{shete2022}%
  \BibitemOpen
  \bibfield  {author} {\bibinfo {author} {\bibfnamefont {K.~P.}\ \bibnamefont
  {Shete}}, \bibinfo {author} {\bibfnamefont {D.~J.}\ \bibnamefont {Boucher}},
  \bibinfo {author} {\bibfnamefont {J.~J.}\ \bibnamefont {Riley}}, \ and\
  \bibinfo {author} {\bibfnamefont {S.~M.}\ \bibnamefont {de~Bruyn~Kops}},\
  }\bibfield  {title} {\enquote {\bibinfo {title} {Effect of viscous-convective
  subrange on passive scalar statistics at high reynolds number},}\ }\href@noop
  {} {\bibfield  {journal} {\bibinfo  {journal} {Physical Review Fluids}\
  }\textbf {\bibinfo {volume} {7}},\ \bibinfo {pages} {024601} (\bibinfo {year}
  {2022})}\BibitemShut {NoStop}%
\bibitem [{\citenamefont {Sreenivasan}\ \emph {et~al.}(1979)\citenamefont
  {Sreenivasan}, \citenamefont {Antonia},\ and\ \citenamefont
  {Britz}}]{sab_79}%
  \BibitemOpen
  \bibfield  {author} {\bibinfo {author} {\bibfnamefont {K.~R.}\ \bibnamefont
  {Sreenivasan}}, \bibinfo {author} {\bibfnamefont {R.~A.}\ \bibnamefont
  {Antonia}}, \ and\ \bibinfo {author} {\bibfnamefont {D.}~\bibnamefont
  {Britz}},\ }\bibfield  {title} {\enquote {\bibinfo {title} {Local isotropy
  and large structures in a heated turbulent jet},}\ }\href@noop {} {\bibfield
  {journal} {\bibinfo  {journal} {J. Fluid Mech.}\ }\textbf {\bibinfo {volume}
  {94}},\ \bibinfo {pages} {745–775} (\bibinfo {year} {1979})}\BibitemShut
  {NoStop}%
\bibitem [{\citenamefont {Warhaft}(2000)}]{ZW00}%
  \BibitemOpen
  \bibfield  {author} {\bibinfo {author} {\bibfnamefont {Z.}~\bibnamefont
  {Warhaft}},\ }\bibfield  {title} {\enquote {\bibinfo {title} {Passive scalars
  in turbulent flows},}\ }\href@noop {} {\bibfield  {journal} {\bibinfo
  {journal} {Annu. Rev. Fluid Mech.}\ }\textbf {\bibinfo {volume} {32}},\
  \bibinfo {pages} {203--240} (\bibinfo {year} {2000})}\BibitemShut {NoStop}%
\bibitem [{\citenamefont {Sreenivasan}(2019)}]{sreeni19}%
  \BibitemOpen
  \bibfield  {author} {\bibinfo {author} {\bibfnamefont {K.~R.}\ \bibnamefont
  {Sreenivasan}},\ }\bibfield  {title} {\enquote {\bibinfo {title} {Turbulent
  mixing: A perspective},}\ }\href@noop {} {\bibfield  {journal} {\bibinfo
  {journal} {Proc.~Natl.~Acad.~Sci.}\ }\textbf {\bibinfo {volume} {116}},\
  \bibinfo {pages} {18175--18183} (\bibinfo {year} {2019})}\BibitemShut
  {NoStop}%
\bibitem [{\citenamefont {Buaria}\ \emph
  {et~al.}(2021{\natexlab{a}})\citenamefont {Buaria}, \citenamefont {Clay},
  \citenamefont {Sreenivasan},\ and\ \citenamefont {Yeung}}]{BCSY21b}%
  \BibitemOpen
  \bibfield  {author} {\bibinfo {author} {\bibfnamefont {D.}~\bibnamefont
  {Buaria}}, \bibinfo {author} {\bibfnamefont {M.~P.}\ \bibnamefont {Clay}},
  \bibinfo {author} {\bibfnamefont {K.~R.}\ \bibnamefont {Sreenivasan}}, \ and\
  \bibinfo {author} {\bibfnamefont {P.~K.}\ \bibnamefont {Yeung}},\ }\bibfield
  {title} {\enquote {\bibinfo {title} {Small-scale isotropy and ramp-cliff
  structures in scalar turbulence},}\ }\href@noop {} {\bibfield  {journal}
  {\bibinfo  {journal} {Phys.~Rev.~Lett.}\ }\textbf {\bibinfo {volume} {126}},\
  \bibinfo {pages} {034504} (\bibinfo {year} {2021}{\natexlab{a}})}\BibitemShut
  {NoStop}%
\bibitem [{Note1()}]{Note1}%
  \BibitemOpen
  \bibinfo {note} {In fact, even in absence of a large-scale mean gradient,
  similar structures are observed in the scalar field \cite
  {watanabe2004}}\BibitemShut {NoStop}%
\bibitem [{\citenamefont {Clay}\ \emph
  {et~al.}(2017{\natexlab{a}})\citenamefont {Clay}, \citenamefont {Buaria},
  \citenamefont {Gotoh},\ and\ \citenamefont {Yeung}}]{clay.cpc1}%
  \BibitemOpen
  \bibfield  {author} {\bibinfo {author} {\bibfnamefont {M.~P.}\ \bibnamefont
  {Clay}}, \bibinfo {author} {\bibfnamefont {D.}~\bibnamefont {Buaria}},
  \bibinfo {author} {\bibfnamefont {T.}~\bibnamefont {Gotoh}}, \ and\ \bibinfo
  {author} {\bibfnamefont {P.~K.}\ \bibnamefont {Yeung}},\ }\bibfield  {title}
  {\enquote {\bibinfo {title} {A dual communicator and dual grid-resolution
  algorithm for petascale simulations of turbulent mixing at high {Schmidt}
  number},}\ }\href@noop {} {\bibfield  {journal} {\bibinfo  {journal} {Comput.
  Phys. Commun.}\ }\textbf {\bibinfo {volume} {219}},\ \bibinfo {pages}
  {313--328} (\bibinfo {year} {2017}{\natexlab{a}})}\BibitemShut {NoStop}%
\bibitem [{\citenamefont {Clay}\ \emph {et~al.}(2018)\citenamefont {Clay},
  \citenamefont {Buaria}, \citenamefont {Yeung},\ and\ \citenamefont
  {Gotoh}}]{clay.cpc2}%
  \BibitemOpen
  \bibfield  {author} {\bibinfo {author} {\bibfnamefont {M.~P.}\ \bibnamefont
  {Clay}}, \bibinfo {author} {\bibfnamefont {D.}~\bibnamefont {Buaria}},
  \bibinfo {author} {\bibfnamefont {P.~K.}\ \bibnamefont {Yeung}}, \ and\
  \bibinfo {author} {\bibfnamefont {T.}~\bibnamefont {Gotoh}},\ }\bibfield
  {title} {\enquote {\bibinfo {title} {{GPU} acceleration of a petascale
  application for turbulent mixing at high {Schmidt} number using {OpenMP}
  4.5},}\ }\href@noop {} {\bibfield  {journal} {\bibinfo  {journal} {Comput.
  Phys. Commun.}\ }\textbf {\bibinfo {volume} {228}},\ \bibinfo {pages}
  {100--114} (\bibinfo {year} {2018})}\BibitemShut {NoStop}%
\bibitem [{\citenamefont {Clay}\ \emph
  {et~al.}(2017{\natexlab{b}})\citenamefont {Clay}, \citenamefont {Buaria},\
  and\ \citenamefont {Yeung}}]{clay.omp}%
  \BibitemOpen
  \bibfield  {author} {\bibinfo {author} {\bibfnamefont {M.~P.}\ \bibnamefont
  {Clay}}, \bibinfo {author} {\bibfnamefont {D.}~\bibnamefont {Buaria}}, \ and\
  \bibinfo {author} {\bibfnamefont {P.~K.}\ \bibnamefont {Yeung}},\ }\bibfield
  {title} {\enquote {\bibinfo {title} {Improving scalability and accelerating
  petascale turbulence simulation using {OpenMP}},}\ }in\ \href@noop {} {\emph
  {\bibinfo {booktitle} {Proceedings of {OpenMP} Conference}}}\ (\bibinfo
  {address} {Stony Brook University, NY},\ \bibinfo {year} {2017})\BibitemShut
  {NoStop}%
\bibitem [{\citenamefont {Buaria}\ \emph
  {et~al.}(2021{\natexlab{b}})\citenamefont {Buaria}, \citenamefont {Clay},
  \citenamefont {Sreenivasan},\ and\ \citenamefont {Yeung}}]{BCSY21a}%
  \BibitemOpen
  \bibfield  {author} {\bibinfo {author} {\bibfnamefont {D.}~\bibnamefont
  {Buaria}}, \bibinfo {author} {\bibfnamefont {M.~P.}\ \bibnamefont {Clay}},
  \bibinfo {author} {\bibfnamefont {K.~R.}\ \bibnamefont {Sreenivasan}}, \ and\
  \bibinfo {author} {\bibfnamefont {P.~K.}\ \bibnamefont {Yeung}},\ }\bibfield
  {title} {\enquote {\bibinfo {title} {Turbulence is an ineffective mixer when
  schmidt numbers are large},}\ }\href@noop {} {\bibfield  {journal} {\bibinfo
  {journal} {Phys.~Rev.~Lett.}\ }\textbf {\bibinfo {volume} {126}},\ \bibinfo
  {pages} {074501} (\bibinfo {year} {2021}{\natexlab{b}})}\BibitemShut
  {NoStop}%
\bibitem [{\citenamefont {Borgas}\ \emph {et~al.}(2004)\citenamefont {Borgas},
  \citenamefont {Sawford}, \citenamefont {Xu}, \citenamefont {Donzis},\ and\
  \citenamefont {Yeung}}]{BSXDY}%
  \BibitemOpen
  \bibfield  {author} {\bibinfo {author} {\bibfnamefont {M.~S.}\ \bibnamefont
  {Borgas}}, \bibinfo {author} {\bibfnamefont {B.~L.}\ \bibnamefont {Sawford}},
  \bibinfo {author} {\bibfnamefont {S.}~\bibnamefont {Xu}}, \bibinfo {author}
  {\bibfnamefont {D.~A.}\ \bibnamefont {Donzis}}, \ and\ \bibinfo {author}
  {\bibfnamefont {P.~K.}\ \bibnamefont {Yeung}},\ }\bibfield  {title} {\enquote
  {\bibinfo {title} {{High Schmidt number scalars in turbulence: structure
  functions and Lagrangian theory}},}\ }\href@noop {} {\ \textbf {\bibinfo
  {volume} {16}},\ \bibinfo {pages} {3888--3899} (\bibinfo {year}
  {2004})}\BibitemShut {NoStop}%
\bibitem [{\citenamefont {Buaria}\ \emph {et~al.}(2016)\citenamefont {Buaria},
  \citenamefont {Yeung},\ and\ \citenamefont {Sawford}}]{BYS.2016}%
  \BibitemOpen
  \bibfield  {author} {\bibinfo {author} {\bibfnamefont {D.}~\bibnamefont
  {Buaria}}, \bibinfo {author} {\bibfnamefont {P.~K.}\ \bibnamefont {Yeung}}, \
  and\ \bibinfo {author} {\bibfnamefont {B.~L.}\ \bibnamefont {Sawford}},\
  }\bibfield  {title} {\enquote {\bibinfo {title} {{A Lagrangian} study of
  turbulent mixing: forward and backward dispersion of molecular trajectories
  in isotropic turbulence},}\ }\href@noop {} {\bibfield  {journal} {\bibinfo
  {journal} {J.~Fluid Mech.}\ }\textbf {\bibinfo {volume} {{799}}},\ \bibinfo
  {pages} {{352--382}} (\bibinfo {year} {2016})}\BibitemShut {NoStop}%
\bibitem [{\citenamefont {Yasuda}\ \emph {et~al.}(2020)\citenamefont {Yasuda},
  \citenamefont {Gotoh}, \citenamefont {Watanabe},\ and\ \citenamefont
  {Saito}}]{yasuda20}%
  \BibitemOpen
  \bibfield  {author} {\bibinfo {author} {\bibfnamefont {T.}~\bibnamefont
  {Yasuda}}, \bibinfo {author} {\bibfnamefont {T.}~\bibnamefont {Gotoh}},
  \bibinfo {author} {\bibfnamefont {T.}~\bibnamefont {Watanabe}}, \ and\
  \bibinfo {author} {\bibfnamefont {I.}~\bibnamefont {Saito}},\ }\bibfield
  {title} {\enquote {\bibinfo {title} {P\'eclet-number dependence of
  small-scale anisotropy of passive scalar fluctuations under a uniform mean
  gradient in isotropic turbulence},}\ }\href@noop {} {\bibfield  {journal}
  {\bibinfo  {journal} {J.~Fluid Mech.}\ }\textbf {\bibinfo {volume} {898}},\
  \bibinfo {pages} {A4} (\bibinfo {year} {2020})}\BibitemShut {NoStop}%
\bibitem [{\citenamefont {Kolmogorov}(1962)}]{K62}%
  \BibitemOpen
  \bibfield  {author} {\bibinfo {author} {\bibfnamefont {A.~N.}\ \bibnamefont
  {Kolmogorov}},\ }\bibfield  {title} {\enquote {\bibinfo {title} {A refinement
  of previous hypotheses concerning the local structure of turbulence in a
  viscous incompressible fluid at high {Reynolds} number},}\ }\href@noop {}
  {\bibfield  {journal} {\bibinfo  {journal} {J.~Fluid Mech.}\ }\textbf
  {\bibinfo {volume} {13}},\ \bibinfo {pages} {82--85} (\bibinfo {year}
  {1962})}\BibitemShut {NoStop}%
\bibitem [{\citenamefont {Prasad}\ \emph {et~al.}(1988)\citenamefont {Prasad},
  \citenamefont {Meneveau},\ and\ \citenamefont {Sreenivasan}}]{prasad88}%
  \BibitemOpen
  \bibfield  {author} {\bibinfo {author} {\bibfnamefont {R.~R.}\ \bibnamefont
  {Prasad}}, \bibinfo {author} {\bibfnamefont {C.}~\bibnamefont {Meneveau}}, \
  and\ \bibinfo {author} {\bibfnamefont {K.~R.}\ \bibnamefont {Sreenivasan}},\
  }\bibfield  {title} {\enquote {\bibinfo {title} {Multifractal nature of the
  dissipation field of passive scalars in fully turbulent flows},}\ }\href@noop
  {} {\bibfield  {journal} {\bibinfo  {journal} {Phys.~Rev.~Lett.}\ }\textbf
  {\bibinfo {volume} {61}},\ \bibinfo {pages} {74} (\bibinfo {year}
  {1988})}\BibitemShut {NoStop}%
\bibitem [{\citenamefont {Sreenivasan}\ and\ \citenamefont
  {Kailasnath}(1993)}]{sreeni93}%
  \BibitemOpen
  \bibfield  {author} {\bibinfo {author} {\bibfnamefont {K.~R.}\ \bibnamefont
  {Sreenivasan}}\ and\ \bibinfo {author} {\bibfnamefont {P.}~\bibnamefont
  {Kailasnath}},\ }\bibfield  {title} {\enquote {\bibinfo {title} {An update on
  the intermittency exponent in turbulence},}\ }\href@noop {} {\bibfield
  {journal} {\bibinfo  {journal} {Phys. Fluids A: Fluid Dynamics}\ }\textbf
  {\bibinfo {volume} {5}},\ \bibinfo {pages} {512--514} (\bibinfo {year}
  {1993})}\BibitemShut {NoStop}%
\bibitem [{\citenamefont {Mydlarski}\ and\ \citenamefont
  {Warhaft}(1998)}]{MW98}%
  \BibitemOpen
  \bibfield  {author} {\bibinfo {author} {\bibfnamefont {L.}~\bibnamefont
  {Mydlarski}}\ and\ \bibinfo {author} {\bibfnamefont {Z.}~\bibnamefont
  {Warhaft}},\ }\bibfield  {title} {\enquote {\bibinfo {title} {Passive scalar
  statistics in high-{P}\'{e}clet-number grid turbulence},}\ }\href@noop {}
  {\bibfield  {journal} {\bibinfo  {journal} {J. Fluid Mech.}\ }\textbf
  {\bibinfo {volume} {358}},\ \bibinfo {pages} {135--175} (\bibinfo {year}
  {1998})}\BibitemShut {NoStop}%
\bibitem [{\citenamefont {Buaria}\ and\ \citenamefont
  {Sreenivasan}(2022)}]{BS2022}%
  \BibitemOpen
  \bibfield  {author} {\bibinfo {author} {\bibfnamefont {D.}~\bibnamefont
  {Buaria}}\ and\ \bibinfo {author} {\bibfnamefont {K.~R.}\ \bibnamefont
  {Sreenivasan}},\ }\bibfield  {title} {\enquote {\bibinfo {title}
  {Intermittency of turbulent velocity and scalar fields using
  three-dimensional local averaging},}\ }\href@noop {} {\bibfield  {journal}
  {\bibinfo  {journal} {Phys.~Rev.~Fluids}\ }\textbf {\bibinfo {volume} {7}},\
  \bibinfo {pages} {L072601} (\bibinfo {year} {2022})}\BibitemShut {NoStop}%
\bibitem [{\citenamefont {Schumacher}\ \emph {et~al.}(2005)\citenamefont
  {Schumacher}, \citenamefont {Sreenivasan},\ and\ \citenamefont
  {Yeung}}]{schum2005}%
  \BibitemOpen
  \bibfield  {author} {\bibinfo {author} {\bibfnamefont {J.}~\bibnamefont
  {Schumacher}}, \bibinfo {author} {\bibfnamefont {K.~R}\ \bibnamefont
  {Sreenivasan}}, \ and\ \bibinfo {author} {\bibfnamefont {P.~K.}\ \bibnamefont
  {Yeung}},\ }\bibfield  {title} {\enquote {\bibinfo {title} {Very fine
  structures in scalar mixing},}\ }\href@noop {} {\bibfield  {journal}
  {\bibinfo  {journal} {J.~Fluid Mech.}\ }\textbf {\bibinfo {volume} {531}},\
  \bibinfo {pages} {113} (\bibinfo {year} {2005})}\BibitemShut {NoStop}%
\bibitem [{\citenamefont {Chen}\ \emph {et~al.}(1997)\citenamefont {Chen},
  \citenamefont {Sreenivasan},\ and\ \citenamefont {Nelkin}}]{chen97}%
  \BibitemOpen
  \bibfield  {author} {\bibinfo {author} {\bibfnamefont {S.}~\bibnamefont
  {Chen}}, \bibinfo {author} {\bibfnamefont {K.~R.}\ \bibnamefont
  {Sreenivasan}}, \ and\ \bibinfo {author} {\bibfnamefont {M.}~\bibnamefont
  {Nelkin}},\ }\bibfield  {title} {\enquote {\bibinfo {title} {Inertial range
  scalings of dissipation and enstrophy in isotropic turbulence},}\ }\href@noop
  {} {\bibfield  {journal} {\bibinfo  {journal} {Phys.~Rev.~Lett.}\ }\textbf
  {\bibinfo {volume} {79}},\ \bibinfo {pages} {1253} (\bibinfo {year}
  {1997})}\BibitemShut {NoStop}%
\bibitem [{\citenamefont {Buaria}\ \emph {et~al.}(2019)\citenamefont {Buaria},
  \citenamefont {Pumir}, \citenamefont {Bodenschatz},\ and\ \citenamefont
  {Yeung}}]{BPBY2019}%
  \BibitemOpen
  \bibfield  {author} {\bibinfo {author} {\bibfnamefont {D.}~\bibnamefont
  {Buaria}}, \bibinfo {author} {\bibfnamefont {A.}~\bibnamefont {Pumir}},
  \bibinfo {author} {\bibfnamefont {E.}~\bibnamefont {Bodenschatz}}, \ and\
  \bibinfo {author} {\bibfnamefont {P.~K.}\ \bibnamefont {Yeung}},\ }\bibfield
  {title} {\enquote {\bibinfo {title} {Extreme velocity gradients in turbulent
  flows},}\ }\href@noop {} {\bibfield  {journal} {\bibinfo  {journal} {New
  J.~Phys.}\ }\textbf {\bibinfo {volume} {21}},\ \bibinfo {pages} {043004}
  (\bibinfo {year} {2019})}\BibitemShut {NoStop}%
\bibitem [{\citenamefont {Buaria}\ and\ \citenamefont {Pumir}(2022)}]{BP2022}%
  \BibitemOpen
  \bibfield  {author} {\bibinfo {author} {\bibfnamefont {D.}~\bibnamefont
  {Buaria}}\ and\ \bibinfo {author} {\bibfnamefont {A.}~\bibnamefont {Pumir}},\
  }\bibfield  {title} {\enquote {\bibinfo {title} {Vorticity-strain rate
  dynamics and the smallest scales of turbulence},}\ }\href@noop {} {\bibfield
  {journal} {\bibinfo  {journal} {Phys.~Rev.~Lett.}\ }\textbf {\bibinfo
  {volume} {128}},\ \bibinfo {pages} {094501} (\bibinfo {year}
  {2022})}\BibitemShut {NoStop}%
\bibitem [{\citenamefont {Ashurst}\ \emph {et~al.}(1987)\citenamefont
  {Ashurst}, \citenamefont {Kerstein}, \citenamefont {Kerr},\ and\
  \citenamefont {Gibson}}]{Ashurst87}%
  \BibitemOpen
  \bibfield  {author} {\bibinfo {author} {\bibfnamefont {W.~T.}\ \bibnamefont
  {Ashurst}}, \bibinfo {author} {\bibfnamefont {A.~R.}\ \bibnamefont
  {Kerstein}}, \bibinfo {author} {\bibfnamefont {R.~M.}\ \bibnamefont {Kerr}},
  \ and\ \bibinfo {author} {\bibfnamefont {C.~H.}\ \bibnamefont {Gibson}},\
  }\bibfield  {title} {\enquote {\bibinfo {title} {Alignment of vorticity and
  scalar gradient with strain rate in simulated {Navier-Stokes} turbulence},}\
  }\href@noop {} {\bibfield  {journal} {\bibinfo  {journal} {Phys. Fluids}\
  }\textbf {\bibinfo {volume} {30}},\ \bibinfo {pages} {2343--2353} (\bibinfo
  {year} {1987})}\BibitemShut {NoStop}%
\bibitem [{\citenamefont {Buaria}\ \emph {et~al.}(2020)\citenamefont {Buaria},
  \citenamefont {Bodenschatz},\ and\ \citenamefont {Pumir}}]{BBP2020}%
  \BibitemOpen
  \bibfield  {author} {\bibinfo {author} {\bibfnamefont {D.}~\bibnamefont
  {Buaria}}, \bibinfo {author} {\bibfnamefont {E.}~\bibnamefont {Bodenschatz}},
  \ and\ \bibinfo {author} {\bibfnamefont {A.}~\bibnamefont {Pumir}},\
  }\bibfield  {title} {\enquote {\bibinfo {title} {Vortex stretching and
  enstrophy production in high {Reynolds} number turbulence},}\ }\href@noop {}
  {\bibfield  {journal} {\bibinfo  {journal} {Phys.~Rev.~Fluids}\ }\textbf
  {\bibinfo {volume} {5}},\ \bibinfo {pages} {104602} (\bibinfo {year}
  {2020})}\BibitemShut {NoStop}%
\bibitem [{\citenamefont {Sreenivasan}\ and\ \citenamefont
  {Antonia}(1997)}]{SA97}%
  \BibitemOpen
  \bibfield  {author} {\bibinfo {author} {\bibfnamefont {K.~R.}\ \bibnamefont
  {Sreenivasan}}\ and\ \bibinfo {author} {\bibfnamefont {R.~A.}\ \bibnamefont
  {Antonia}},\ }\bibfield  {title} {\enquote {\bibinfo {title} {The
  phenomenology of small-scale turbulence},}\ }\href@noop {} {\bibfield
  {journal} {\bibinfo  {journal} {Annu.~Rev.~Fluid~Mech.}\ }\textbf {\bibinfo
  {volume} {29}},\ \bibinfo {pages} {435--77} (\bibinfo {year}
  {1997})}\BibitemShut {NoStop}%
\bibitem [{\citenamefont {Watanabe}\ and\ \citenamefont
  {Gotoh}(2004)}]{watanabe2004}%
  \BibitemOpen
  \bibfield  {author} {\bibinfo {author} {\bibfnamefont {T.}~\bibnamefont
  {Watanabe}}\ and\ \bibinfo {author} {\bibfnamefont {T.}~\bibnamefont
  {Gotoh}},\ }\bibfield  {title} {\enquote {\bibinfo {title} {Statistics of a
  passive scalar in homogeneous turbulence},}\ }\href@noop {} {\bibfield
  {journal} {\bibinfo  {journal} {New J.~Phys.}\ }\textbf {\bibinfo {volume}
  {6}},\ \bibinfo {pages} {40} (\bibinfo {year} {2004})}\BibitemShut {NoStop}%
\end{thebibliography}

%merlin.mbs apsrev4-1.bst 2010-07-25 4.21a (PWD, AO, DPC) hacked
%Control: key (0)
%Control: author (0) dotless jnrlst
%Control: editor formatted (1) identically to author
%Control: production of article title (0) allowed
%Control: page (1) range
%Control: year (0) verbatim
%Control: production of eprint (0) enabled
%

\end{document}